# New pixelized Micromegas detector with low discharge rate for the COMPASS experiment

Damien Neyret[1], Philippe Abbon, Marc Anfreville, Yann Bedfer, Etienne Burtin, Christophe Coquelet, Nicole d'Hose, Daniel Desforge, Arnaud Giganon, Didier Jourde, Fabienne Kunne, Alain Magnon, Nour Makke, Claude Marchand, Bernard Paul, Stéphane Platchkov, Florian Thibaud, Michel Usseglio, and Maxence Vandenbroucke

*CEA Saclay DSM IRFU,*
  *91191 Gif sur Yvette Cedex, France*
  *E-mail*: damien.neyret@cea.fr

ABSTRACT: New Micromegas (*Micro-me*sh *gas*eous detectors) are being developed in view of the future physics projects planned by the COMPASS collaboration at CERN. Several major upgrades compared to present detectors are being studied: detectors standing five times higher luminosity with hadron beams, detection of beam particles (flux up to a few hundred of kHz/mm², 10 times larger than for the present Micromegas detectors) with pixelized read-out in the central part, light and integrated electronics, and improved robustness. Two solutions for a reduction of the impact of discharges have been studied, with Micromegas detectors using resistive layers and using an additional GEM foil. The performance of such detectors has been measured during beam test periods. A large size prototype with nominal active area and pixelized read-out has been produced and installed in the COMPASS spectrometer in 2010. In 2011 prototypes featuring an additional GEM foil, as well as a resistive prototype, were tested in similar conditions and preliminary results from those detectors are very promising. We present here the project and report on its status, in particular the performance of large size prototypes with an additional GEM foil.

---

[1] Corresponding author.

# Contents

## 1. The COMPASS experiment and the Micromegas detectors

The COMPASS experiment [1] is dedicated to the study of the spin structure of the nucleon and the spectroscopy of hadrons. It takes advantage of the secondary muon and hadron beams delivered by the M2 beam line at the SPS (Super Proton Synchrotron) accelerator at CERN, with an energy range from 100 to above 200 GeV and with beam intensities reaching $10^8$ part/s. The COMPASS spectrometer consists of a fixed target and a two-stage spectrometer for the detection and identification of particles at low and high momenta at high flux.

Since the beginning of the COMPASS data taking in 2002, 12 Micromegas detectors are installed at the downstream edge of the fixed target to detect particles scattered at low angles. Micromegas detectors are fast gaseous detectors using a micro-mesh electrode separating a thick low-field ionization area, where the gas is ionized by the charged particles, and a high-field amplification gap where the electrons produced in the ionization gap are amplified and read by micro-strips. As the amplification gap is very thin (in the order of 100 µm) the ions are quickly neutralized. The produced signals are then much shorter, in the order of 100 ns, than standard multi-wires chambers.

These Micromegas detectors [2] were developed by the CEA Saclay in order to fit the requirements of the COMPASS experiment to stand a flux up to 500 kHz/cm², with a spatial resolution better than 100 µm and low material budget. The amplification gap is 100 µm thick, with an ionization gap of 5 mm [3]. The detectors are filled with a gas mixture of Neon + 10% ethane + 10% $CF_4$. The drift and mesh electrodes are made of 5 µm copper meshes. The voltage applied on the amplification gap is around 400 to 420 V, giving a gain between 3000 to 6000. The active area of each detector covers 40x40 cm², with a blind disk of 5 cm diameter on the path of the beam. These detectors measure particle positions in one dimension, the 12 detectors being grouped by stations of 4 detectors each, covering X, Y, U and V (±45°) coordinates. The micro-strips are extended outside the active area by 30 cm to shift the electronic cards out of the spectrometer acceptance.

The detection efficiency is better than 98% at low particle flux, while at high flux it decreases to 96% due to the electronics occupancy, with a counting rate which can reach 150 kHz/channel. In nominal high flux conditions the spatial resolution is between 90 and 100 µm, except for detectors close to the polarized target and dipole magnetic fringe fields (up to 1 T), where the resolution is 10% larger due to the Lorentz angle effect. The discharge rate has been minimized by using a light gas (Ne + 10% $C_2H_6$ + 10% $CF_4$) and also by running the detectors at moderate gain thanks to the low noise electronics using SFE16 chips [4]. With a high flux muon beam ($4.10^7$ muon/s on a 1.20 m $^6$LiD solid target), the discharge rate per detector is lower than 0.02 discharge/s. With a hadron beam the discharge rate reaches 0.04 to 0.1 discharge/s with a beam of $10^7$ hadron/s. This rather high discharge rate is tolerable but prevents any future increase of the hadron beam flux. Higher discharge rates would reduce the detector efficiency and might deteriorate the hardware.

## 2. Project for a new pixelized Micromegas detector

The COMPASS collaboration is completing in 2011 its initial physics program [5]. For the long term future (2012 and beyond) a new proposal has been submitted [6]. New physics objectives are the study of Generalized Parton Distributions via Deeply Virtual Compton Scattering, transversity via the study of the Drell-Yan reaction and further measurements on hadron spectroscopy.

The perspective to use Micromegas detectors during many more years and at a higher flux, lead us to take the opportunity to improve them. Two major objectives have been defined: the new detectors must stand more than 5 times higher hadron beam flux, and they must be active in the central region where the beam is crossing in order to replace present thick scintillating fiber detectors.

The goal of the present R&D project is to develop new Micromegas detectors which will fulfill these requirements, with 10 to 100 lower discharge rate compared to present detectors, in order to take into account the activation of the central area, and with a pixelized read-out in the beam area. A light integrated read-out electronics using APV25-S1 chip [7] is developed, and the new "bulk" Micromegas technology [8] has been chosen to build those detectors, as it produces more reliable chambers.

### 2.1 Discharge impact reduction

Several solutions to reduce the impact of the discharges have been considered. These solutions are either based on resistive layers applied on the anode strips, or on an additional GEM foil above the Micromegas mesh.

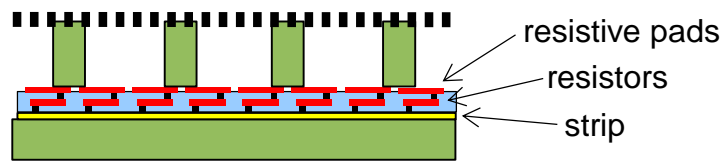

*Figure 1: Scheme of buried resistor resistive detector*

The role of the resistive layer is to minimize the impact of the discharges by limiting the current to the strips and thus the duration of the discharges. The capacitance involved in the discharge process would also be limited by the resistive layer. Several sorts of coating were considered: resistive foil on isolating layer, resistive coating in contact with the strips, segmented resistive coating, etc... Resistive coatings are studied since a few years, in particular by sLHC and ILC CEA Saclay groups, as well as other groups, and interesting results were already presented [9]. Since 2009 we also studied several schemes of resistive Micromegas detectors, but none of them showed performance adapted to our needs, in particular in term of reliability or of spatial resolution, excepted one scheme. This late scheme (Figure 1) was proposed beginning of 2010 by Rui de Oliveira [10]. According to this scheme, the anode strips are covered by resistive pads connected to the strips by intermediate resistors placed horizontally. Signals coming from crossing particles are transmitted to the strips by capacitive coupling, while the charges are evacuated by the intermediate resistor. When a discharge occurs, the electric potential of the pad rises quickly to reach the mesh potential value, stopping the discharge. The internal resistor in this scheme is long enough to avoid any breakdown or damage from the discharge.

The other solution also studied is to use an additional GEM foil which will preamplify the primary electrons. This foil allows reducing the gain of the micro-mesh stage and thus reducing the discharge probability for the same total gain. Results from both solutions are presented extensively in these proceedings by M. Vandenbroucke [11] and S. Procureur [12].

### 2.2 Design of the large size pixelized detector

The particle flux in the center of the COMPASS Micromegas detectors is expected to be very high, of the order of 100 to 200 kHz/mm², since the incoming beam goes through the detectors here. A read-out with strips would lead to hit rates of the order of 500 kHz/channel, generating an inefficiency due to the electronics occupancy larger than 10%.

To decrease the hit rate to about 200 kHz/channel, a pixelized read-out with pixel surfaces of about 1 and 2.5 mm² in the central area has been chosen. To keep a good spatial resolution these pixels are rectangular and parallel to the strips, with a size of 2.5x0.4 mm² in the center and 6.25x0.4 mm² at larger angles (see Figure 2). The remaining of the 40x40 cm² active area is covered by 20 cm long strips with the same 400 μm pitch in the center of the detector, and 40 cm long strips with 480 μm pitch on the edges. The number of electronics channels to read the pixels is 1280, in addition to the 1280 strip channels.

The material budget of the pixelized prototypes is similar to the present COMPASS Micromegas detectors, with a radiation length of around 0.32% versus 0.287% of $X_0$. The difference is due to the use of a woven stainless steel mesh instead of a 5 μm thick copper mesh.

### 2.3 Read-out electronics with APV chips

The present read-out electronics using SFE16 amplifiers/discriminators in association with F1 TDC chips is working well since the beginning of the COMPASS experiment. However the space taken by the electronic cards is quite important as each card reads only a small number of channels (64 for a F1 board). The power consumption is also important, around 500 W per detector, requiring a high cooling power.

As the new detectors will have more than twice more channels, it is not desirable to use the same electronics, due to the lack of space and the high heat production. A more compact and integrated electronics is being studied, using APV chips which are 128-channel amplifiers and analog multiplexers. The characteristics of the amplifier are tunable using configuration registers, and its timing constants can be adapted to the characteristics of the detector. The analog values are sampled at 40 MHz and when an event is triggered, three samples from each channel are multiplexed and sent to a digitization card. This card features a 10-bit flash ADC for each connected APV, and a FPGA which applies several algorithms: pedestal subtractions, common mode noise correction, and zero suppression.

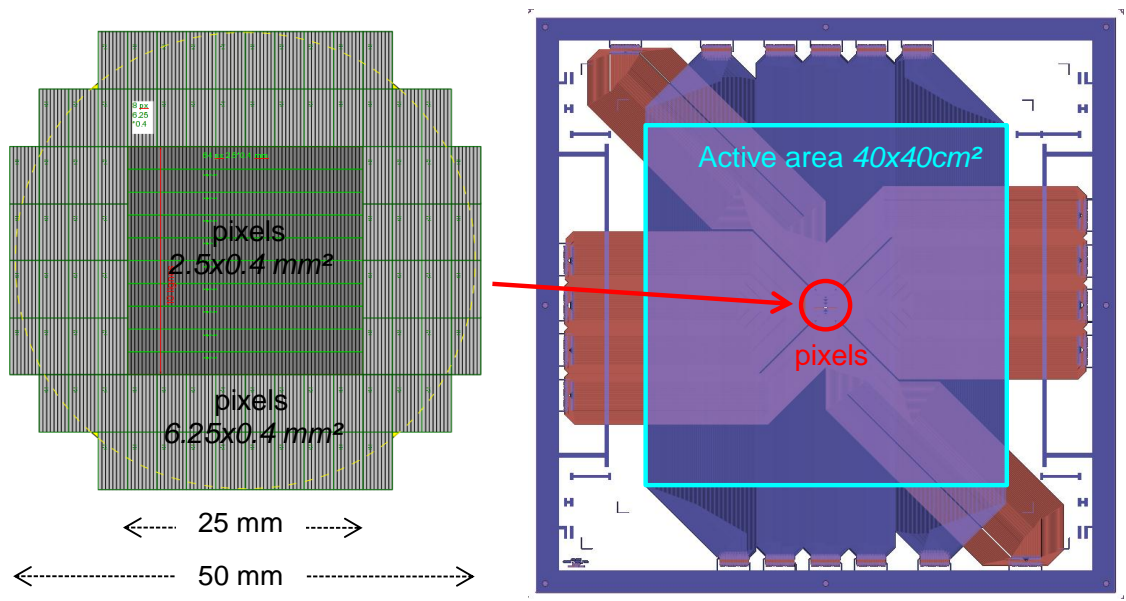

*Figure 2: Sketch of large pixelized Micromegas detector (right). Zoom of the pixel area (left)*

APV electronic read-out is already used in the COMPASS experiment. It was developed by the E18 group of the Technische Universität of Munich (TUM) for the GEM detectors [13], the Silicon detectors [14] and the new pixel GEM detectors [15]. A common project with the CEA Saclay has also permitted to develop a fast APV read-out for the multi-wire proportional chambers of the RICH detector [16]. Front-end electronic cards have been adapted to the Micromegas detectors, in particular the protection and decoupling circuit in front of the APV.

An important feature of this electronics is its high density: an APV card reads 128 channels, and an ADC board, connected to 16 APV cards, reads 2048 channels. Only 20 APV cards and 2 ADC cards are needed.

### 2.4 Improving robustness with bulk technology

Some of the old Micromegas detectors exhibited a few mechanical issues (gluing defaults, tightening of copper mesh, impurities below the mesh) concerning the board and the thin copper meshes. The board, being built from 100 μm thin epoxy layers, was glued on a 5 mm thick honeycomb, with the 5 μm copper mesh mounted on a frame.

The bulk technology is a way to improve the robustness of the detectors. A woven stainless steel mesh is laminated to the board between two photosensitive coverlays, and an UV exposure is applied with appropriate masks to draw pillars on the coverlays. At the end of the process the mesh is completely fixed to the board and is dust tight. It is also less sensitive to the gluing default of the epoxy layer to the honeycomb. Compared to previous detectors using thin copper meshes, bulk detectors are using thicker woven stainless steel meshes with 18 μm wires. Performances of both technologies were compared but the difference in term of gain, efficiencies and discharge probabilities are rather small. The increase of radiation length is also small compared to the other contributions.

## 3. Status of the project

### 3.1 Discharge impact reduction studies

Small 10x10 cm² and 6x10 cm² prototypes of Micromegas detectors have been built and tested in laboratory and beam in 2009 and 2010, in order to test the solutions mentioned in Section 2. In 2010 two prototypes were equipped with a GEM foil placed at 1 and 2 mm above the mesh, while other prototypes were featuring different types of resistive layers. In particular two prototypes built with buried resistors were produced in summer 2010.

These prototypes were tested at CERN SPS with 150 GeV muon and hadron beams in 2009 and 2010, and at the PS with 0.4 to 3 GeV hadron beams in June 2010. Prototypes featuring an additional GEM foil showed spark probabilities reduced by factors 10 to 100 compared to standard detectors at the same gain. Efficiencies and spatial resolution measured with muon beam were very similar to classic bulk Micromegas detectors. These results are presented in detail in the present proceedings, in the contributions of M. Vandenbroucke [11] and S. Procureur [12].

These tests permitted to validate the solution of Micromegas detectors featuring a GEM foil. Resistive detectors were also tested but did not show performances or reliabilities compatible with our needs, except those with the buried resistors. One of these later prototypes showed very promising performances in term of efficiencies and spatial resolution. Discharge amplitudes were also low enough to be not visible to the spark tagging system. However further studies should be done in order to fully validate this solution.

### 3.2 Pixelized Micromegas detectors

A large size pixelized detector has been produced at CERN and CEA Saclay in 2010 and tested in the COMPASS beam in order to evaluate the behavior of such detector in nominal conditions. Two other prototypes featuring a GEM foil 2 mm above the mesh have been produced beginning of 2011 in order to validate this design. Each prototype was successively installed at COMPASS between two of the currently installed Micromegas stations and was exposed to the same conditions. Data analysis of these detectors is in progress, however preliminary results are given here.

A resistive prototype using the buried resistor scheme has been produced in October 2011 by the CERN TE-MPE-EM group and was tested in the same conditions at COMPASS. On-line results showed promising behavior of the prototype, however studies are in progress in order to extract its performances.

#### 3.2.1 Efficiencies

Efficiencies of the detectors were measured using the standard track reconstruction software of the COMPASS experiment. Data from other COMPASS detectors were used to reconstruct tracks corresponding to particles passing through the experimental set-up, then these tracks were compared to the data provided by the prototype. The efficiencies are defined as the probability for a recognized track to correspond to a detector cluster within a rather large route of 2 mm, the effect of background signals being corrected [17].

Preliminary 2D efficiencies of the first prototype with GEM foil are shown in Figure 3 for a gain around 10000 and in the presence of the high intensity COMPASS muon beam ($2.10^7$ muon/s). The performance of the detector is very good, with homogeneous efficiencies above 95% over the active area. The pixel area, zoomed on the right, also shows a good efficiency except for long pixels in the vicinity of the beam. This effect may be due to a too large occupancy of the APV electronics for these channels, or a failure in the data reconstruction of the pixel part. Further studies are ongoing in order to get rid of this effect. In particular data have been taken with different tuning of the APV electronics with shorter shapes of the signals after the amplifier stage. A compromise should be found between the total occupancy of the signal which should be as short as possible in order to reduce the occupancy of the electronics, and the peaking time which should be large enough to integrate as much as possible the signals from the Micromegas detector (around 100 ns).

The efficiency dependence versus the detector gain is shown in Figure 4. A very large efficiency plateau is obtained, starting at a gain smaller than 10000 and reaching an asymptotic efficiency above 95%.

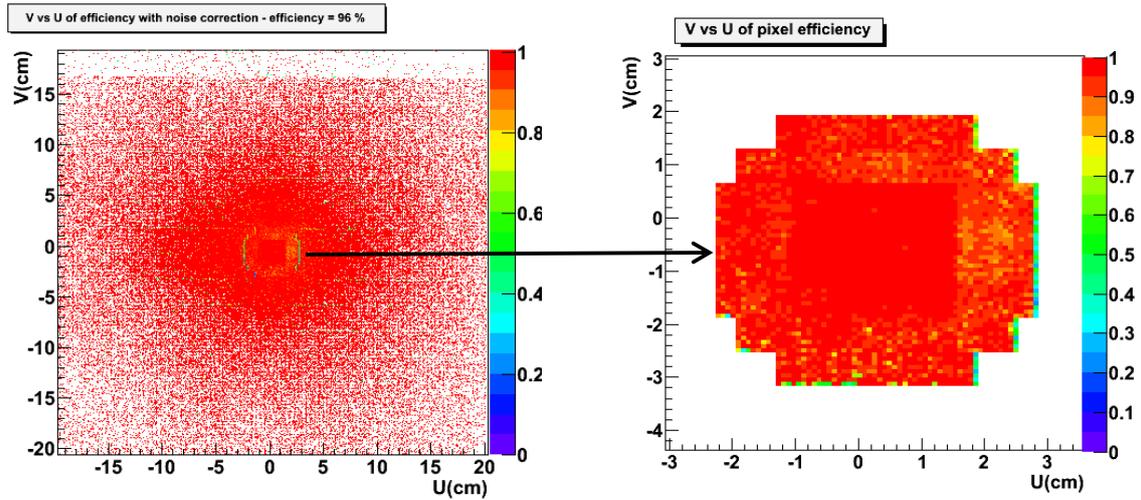

*Figure 3: 2D efficiencies of the first Micromegas with GEM foil prototype for $HV_{mesh}=340V$ and $HV_{GEM}=300V$ (left). Zoom of the pixel area (right).*

### 3.2.2 Spatial resolution

Measurement of the spatial resolution of the prototypes is still in progress. Figure 5 shows preliminary measurements of the residual distribution, which is the difference between the particle position measured by the COMPASS tracking system and the one measured by the prototype. The spatial resolutions have not yet been extracted from the residuals distributions. The left plot shows residual distribution of the first pixelized Micromegas prototype featuring a GEM foil, with pixel area excluded. This distribution can be fitted only with two Gaussian curves, due to a bad position along the beam of the detector taken into account by the reconstruction software. In this case, particles coming from the target and halo particles do not generate the same distributions, the first ones being affected by the additional error (halo trigger was activated for these particular runs in order to populate detector surface far from the center). However the narrow Gaussian curve due to halo particles shows a sigma lower than 95 μm, which is comparable to what one can obtain from the detectors presently installed at COMPASS [18]. The plot on the right side of the figure shows the same residual distribution for the pixel area. The narrow Gaussian curve fitted on that distribution shows a sigma below 135 μm, which is not as good as the strip area, probably due to the high particle flux close to the beam.

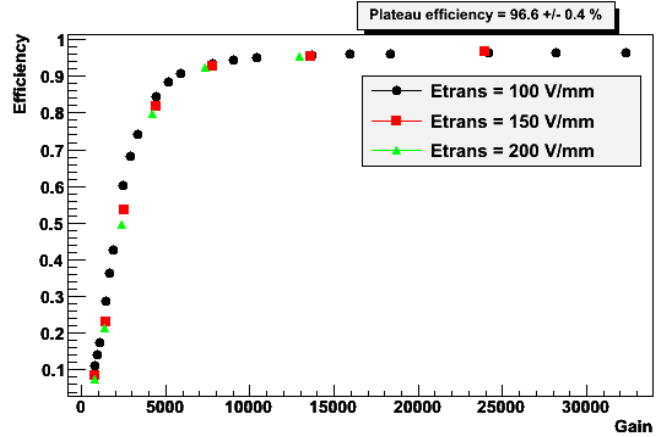

*Figure 4: Efficiency plateau vs detector gain of the first Micromegas with a GEM foil prototype*

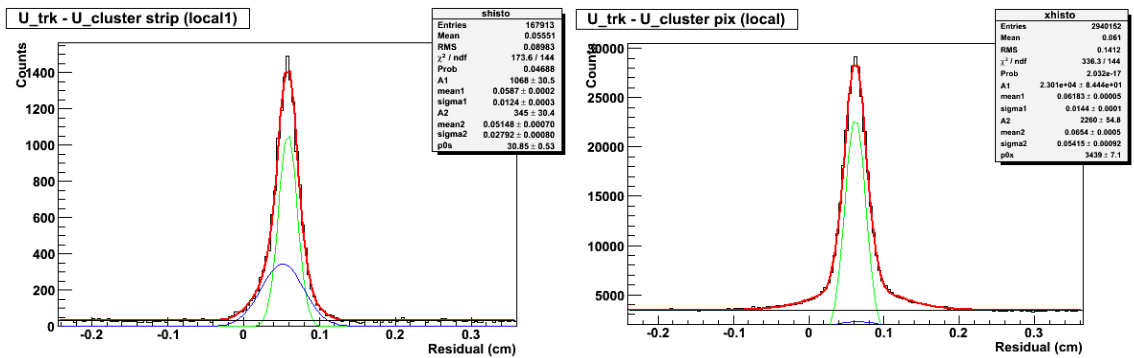

*Figure 5: Spatial residuals (in cm) of the first large size Micromegas detector with a GEM foil, for strip (left) and for pixel (right) areas.*

## 4. Summary and perspective

Large size prototypes with nominal active area and pixelized read-out have been installed and tested at COMPASS in 2011. Two detectors were featuring a GEM foil while a resistive prototype using the buried resistor scheme was tested at the end of the beam period. Analysis of these data is ongoing and performance of these detectors will be measured soon.

In 2012 the COMPASS experiment will take data using a high energy and high flux hadron beam. Both detectors featuring a GEM foil will be installed at COMPASS replacing two of the presently installed Micromegas detectors, and will be fully integrated into the COMPASS spectrometer. The resistive prototype will also be temporarily installed in order to measure its performance with a hadron beam and to fully validate this solution.

In parallel to the COMPASS data taking, an R&D will be done in common between CEA Saclay, the CNRS Subatech and LAPP laboratories, and the CIREA company (PCB producer in France), with the perspective to industrialize the production of large-size bulk Micromegas boards. A grant from the French ANR financing agency has been allocated to this project. Large-size prototypes are expected to be produced in the first half of 2012. The CIREA company will also participate in the development of the production of resistive boards.

The definitive choice of the technology used to reduce the discharge impact will be done in 2012 for a production starting in 2013. New pixelized Micromegas detectors are expected to fully replace the present ones for the COMPASS run of 2014.

## Acknowledgments

We would like to thank the E18 group of the Technische Universität of Munich, and in particular Stephan Paul, Bernhard Ketzer, Igor Konorov and Florian Haas for their great help. We also thank Leszek Ropelewski from CERN GEM group, as well as the CERN TE-MPE-EM laboratory, especially Rui de Oliveira and Olivier Pizzirusso. SPS beam tests have been accomplished in the framework and with the great help of the RD51 collaboration. Part of this work has been financed by the "Physique des 2 Infinis" cluster.